\documentclass{emulateapj}
\shortauthors{Leith, Ng \& Wiltshire}
\shorttitle{Gravitational energy as dark energy}

\begin{document}
\font\sevenrm=cmr7
\def\dd{{\rm d}} \def\ds{\dd s} \def\e{{\rm e}} \def\etal{{\em et al}.}
\def\al{\alpha}\def\be{\beta}\def\ga{\gamma}\def\de{\delta}\def\ep{\epsilon}
\def\et{\eta}\def\th{\theta}\def\ph{\phi}\def\rh{\rho}\def\si{\sigma}

\def\mean#1{{\vphantom{\tilde#1}\bar#1}}
\def\bH{\mean H}\def\OM{\mean\Omega}
\def\OmB{\Omega\Z B}\def\bOmB{{\mean\Omega}\Z B}\def\bnB{\bn\Z B}
\def\gb{\mean\ga}\def\gc{\gb\Z0}\def\etb{\mean\eta}
\def\OMBn{\mean\Omega\Z{B0}}
\def\zdec{z\ns{dec}} \def\OmBn{\Omega\Z{B0}}\def\bOmBn{\mean\Omega\Z{B0}}
\def\kB{k\Z B} \def\Ns#1{\Z{\hbox{\sevenrm #1}}}\def\ze{\,\zeta(3)\,}
\def\DE{\Delta} \def\la{\lambda}
\def\w#1{\,\hbox{#1}} \def\Deriv#1#2#3{{#1#3\over#1#2}}
\def\Der#1#2{{#1\hphantom{#2}\over#1#2}} \def\br{\hfill\break}
\def\ts{t} \def\tc{\tau} \def\Dtc{\mathop{\hbox{$\Der\dd\tc$}}}
\def\Dts{\mathop{\hbox{$\Der\dd\ts$}}}
\def\ns#1{_{\hbox{\sevenrm #1}}} \def\dOM{\dd\Omega^2}
\def\goesas{\mathop{\sim}\limits} \def\twn{\tc\Z0}
\def\Z#1{_{\lower2pt\hbox{$\scriptstyle#1$}}}
\def\X#1{_{\lower2pt\hbox{$\scriptscriptstyle#1$}}}
\def\av{{a\ns{v}\hskip-2pt}} \def\aw{{a\ns{w}\hskip-2.4pt}}
\def\ac{a} \def\an{\ac\Z0} \def\QQ{{\cal Q}} \def\qh{q}
\def\Om{\Omega\Z M} \def\gw{\gb\ns w} \def\tw{\tc}
\def\fw{{f\ns w}} \def\fvi{{f\ns{vi}}} \def\fwi{{f\ns{wi}}}
\def\etw{\eta\ns w} \def\etv{\eta\ns v} \def\rw{r\ns w}
\def\gv{\gb\ns v} \def\fv{{f\ns v}} \def\Hv{H\ns v} \def\Hw{H\ns w}
\def\FF{{\cal F}} \def\FI{\FF\Z I} \def\Fi{\FF\X I}
\def\OMM{\OM\Z M}\def\OMk{\OM_k}\def\OMQ{\OM\Z{\QQ}}\def\OMR{\OM\Z R}
\def\OmM{\Omega\Z M} \def\OmR{\Omega\Z R} \def\OMMn{\OM\Z{M0}}
\def\OmMw{\Omega\Z{M\hbox{\sevenrm w}}} \def\OmMn{\Omega\Z{M0}}
\def\fvf{\left(1-\fv\right)} \def\Hb{\bH\Z{\!0}} \def\Hh{H} \def\Hm{H\Z0}
\def\hri{h_{ri}}
\def\half{\frn12} \def\DD{{\cal D}} \def\bx{{\mathbf x}} \def\Vav{{\cal V}}
\def\frn#1#2{{\textstyle{#1\over#2}}} \def\LL{${\cal L}$}
\def\lsim{\mathop{\hbox{${\lower3.8pt\hbox{$<$}}\atop{\raise0.2pt\hbox{$\sim$}}
$}}} \def\dL{d\Z L} \def\dA{d\Z A} \def\rhcr{\rh\ws{cr}}
\def\kmsMpc{\w{km}\;\w{sec}^{-1}\w{Mpc}^{-1}}
\def\etBg{\et\Z{B\ga}} \def\ab{{\bar a}}
\def\lcdm{(\Lambda\hbox{\sevenrm CDM})}
\def\epi{\epsilon_i}\def\gbi{\gb_i}\def\Omi{\OM_i}\def\OMkn{\OM_{k0}}
\def\fvn{{f\ns{v0}}} \def\Ci{C_\epsilon} \def\te{t_\epsilon}
\def\LCDM{$\Lambda$CDM}
\def\PRL#1{Phys.\ Rev.\ Lett.\ {\bf#1}} \def\PR#1{Phys.\ Rev.\ {\bf#1}}
\def\ApJ#1{ApJ {\bf#1}} \def\PL#1{Phys.\ Lett.\ {\bf#1}}
\def\MNRAS#1{MNRAS {\bf#1}}
\def\GRG#1{Gen.\ Relativ.\ Grav.\ {\bf#1}}
\def\AA#1{Astron.\ Astrophys.\ {\bf#1}}
\def\MNRAS#1{MNRAS {\bf#1}}
\def\GRG#1{Gen.\ Relativ.\ Grav.\ {\bf#1}}
\def\figdmu{Fig.~1} \def\figcont{Fig.~2}

\title{Gravitational energy as dark energy:
concordance of cosmological tests}

\author {{Ben~M.~Leith\altaffilmark{1}}, {S.C.~Cindy~Ng\altaffilmark{2}},
and {David~L.~Wiltshire\altaffilmark{1}}}

\altaffiltext{1}{Department of Physics \& Astronomy,
University of Canterbury, Private Bag 4800, Christchurch 8140, New Zealand}
\altaffiltext{2}{Physics Department, National University of Singapore,
2 Science Drive 3, Singapore 117542}

\begin{abstract}
We provide preliminary quantitative evidence that a new solution to averaging
the observed inhomogeneous structure of matter in the universe
\citep{opus,sol}, may lead to an observationally
viable cosmology without exotic dark energy. We find parameters
which simultaneously satisfy three independent tests: the match to the
angular scale of the sound horizon detected in the cosmic microwave
background anisotropy spectrum; the effective comoving baryon acoustic
oscillation scale detected in galaxy clustering statistics; and type Ia
supernova luminosity distances. Independently of the supernova data,
concordance is obtained for a value of the Hubble constant which agrees with
the measurement of the Hubble Key team of \citet{Sandage}.
Best--fit parameters include a global average Hubble constant
$\Hm=61.7^{+1.2}_{-1.1}\kmsMpc$, a present epoch void volume fraction of
$\fvn=0.76^{+0.12}_{-0.09}$, and an age of the universe of $14.7^{+0.7}_{-0.5}$
billion years as measured by observers in galaxies. The mass ratio of
non--baryonic dark matter to baryonic matter is $3.1^{+2.5}_{-2.4}$,
computed with a baryon--to--photon ratio that concords with primordial
lithium abundances.
\end{abstract}
\maketitle

The apparent acceleration in present cosmic expansion
is usually attributed to a smooth ``dark energy'',
whose nature poses a foundational mystery to physics. Our
standard \LCDM\ cosmology, with a cosmological constant, $\Lambda$, as
dark energy, fits three independent observational tests: type Ia supernovae
(SneIa) luminosity distances; the angular scale of the Doppler
peaks in the spectrum of cosmic microwave background (CMB) temperature
anisotropies; and the baryon acoustic oscillation scale detected in galaxy
clustering statistics. In this {\em Letter} we provide preliminary
evidence that these same tests can all be satisfied in ordinary
general relativity without exotic dark energy, within a model \citep{opus,sol}
which takes a new approach to averaging the observed structure of the
universe, presently dominated by voids.

Recently a number of cosmologists have questioned
whether cosmic acceleration might in fact be
an artifact of replacing the actual observed structure of the universe by
a smooth featureless dust fluid in Einstein's equations. (For a review
see \citet{Brev}.)
The specific solution to the averaging problem we investigate here
\citep{opus} realises cosmic acceleration as an apparent effect
that arises in the decoupling of bound
systems from the global expansion of the universe. In particular, gradients
in the kinetic energy of expansion, and more importantly, in the quasilocal
energy associated with spatial curvature gradients between bound systems
and a volume--average position in freely expanding space, can manifest
themselves in a significant difference in clock rates between the two
locations. This difference is negligible in the early
universe when the assumption of homogeneity is valid, but becomes important
after the transition to void dominance, making apparent acceleration a
phenomenon registered by observers in galaxies at relatively late epochs.

Galaxies and other objects dense enough to be observed at cosmological
distances are bound systems, leading to a selection bias in our sampling
of cosmic clocks. Since
the clock rates within bound systems are closely tied to a universal
{\em finite infinity} scale \citep{fit1,opus}, gross variations in cosmic clock
rates are not directly observable in any observational test yet devised.
However, relative to observers in bound systems an ideal comoving observer
within a void would measure an older age of the universe, and an
isotropic CMB with a lower mean temperature and an angular anisotropy scale
shifted to smaller angles.

A systematic variation in clock rates between bound systems and the volume
average, which we will find to be 38\% at the present epoch,
seems implausible given our familiarity
of large gravitational time dilation effects occurring only for extreme
density contrasts, such as with black holes. However, cosmology presents
a circumstance in which conventional intuition based on static Newtonian
potentials can fail, because spacetime itself is dynamical and the definition
of gravitational energy is extremely subtle. The normalization of clock
rates in bound systems relative to expanding regions can accumulate
significant differences, given that the entire age of the universe
has been available for this to occur.

In this {\em Letter} we find best--fit parameters for the two--scale
{\em fractal bubble} (FB) model \citep{opus,sol}.
The two scales represent voids, and the filaments
and bubble walls which surround them, within which clusters of galaxies are
located. The geometry within finite infinity regions in the bubble walls is
assumed to be spatially flat, but the geometry beyond these regions
is not spatially flat. The relationship between the geometry in galaxies and
the volume--average geometry within our present horizon volume is fixed by
the assumption that the regionally ``locally'' measured expansion is uniform
despite variations in spatial curvature and clock rates. This provides an
implicit resolution of the Sandage--de Vaucouleurs paradox \citep{opus}: the
``locally'' measured or ``bare'' Hubble flow is uniform, but since clock rates
vary it will appear that voids expand faster than walls when referred to
any single set of clocks.

As observers in galaxies, our local average geometry at the
boundary of a finite infinity region is spatially flat, with the metric
\begin{equation}\ds^2\Z{\Fi}=-\dd\tw^2+\aw^2(\tw)\left[\dd\etw^2+
\etw^2\dd\Omega^2\right]\,.
\label{figeom}\end{equation}
Finite infinity regions are contained within filaments and bubble walls.
These walls surround voids, where the metric is not given by (\ref{figeom})
but is negatively curved, with local scale factor $\av$.
The {\em average} geometry is determined by a solution of the Buchert
equations \citep{buch1}, with average
scale factor $\ab^3=\fwi\aw^3+\fvi\av^3$, where $\fvi\ll1$ and $\fwi=1-\fvi$
are the respective initial void and wall volume fractions at last
scattering, when the assumption of homogeneity is justified by the evidence
of the CMB and the Copernican principle. It takes the form
\begin{equation}
\ds^2=-\dd\ts^2+\ab^2(\ts)\,\dd\etb^2+A(\etb,\ts)\,\dOM,
\label{avgeom}
\end{equation}
where the area function $A$ is defined by a horizon-volume average
\citep{opus}. The time--parameter $\ts$ differs from the wall--time $\tw$
of (\ref{figeom}) by the mean lapse function $\dd\ts=\gb(\tw)\,\dd\tw$.
The geometry (\ref{avgeom}) does not match the local geometry in either the
walls or void centres.

When the geometry (\ref{figeom}) is related to the average
geometry (\ref{avgeom}) by conformal matching of radial null geodesics
it may be rewritten
\begin{equation} \ds^2\Z{\Fi}=
-\dd\tw^2+{\ab^2\over\gb^2}\left[\dd\etb^2+\rw^2(\etb,\tw)\,\dOM\right]
\label{wgeom}
\end{equation}
where
$\rw\equiv\gb\fvf^{1/3}\fwi^{-1/3}\etw(\etb,\tw)$. Two
sets of cosmological parameters are relevant: those
relative to an ideal observer at the volume--average position in freely
expanding space using the metric (\ref{avgeom}), and conventional
{\em dressed parameters} using the metric (\ref{wgeom}). The conventional
metric (\ref{wgeom}) arises in our attempt to fit a single global metric
(\ref{figeom}) to the universe with the assumption that average spatial
curvature and local clock rates everywhere are identical to our own,
which is no longer true. One consequence is that the dressed
matter density parameter, $\OmM$, differs from the bare volume--average
density parameter, $\OMM$, according to $\OmM=\gb^3\OMM$.

The conventional dressed Hubble parameter, $H$, of the metric (\ref{wgeom})
differs from the bare Hubble parameter, $\bH$, of (\ref{avgeom})
according to
\begin{equation}
\Hh=\gb\bH-\Dts\gb=\gb\bH-\gb^{-1}\Dtc\gb\,.
\label{42}
\end{equation}
Since the bare Hubble parameter characterizes the uniform ``locally measured''
Hubble flow, its present value coincides with the value of the Hubble constant
that observers in galaxies would obtain for measurements
averaged solely within the plane of an ideal local bubble wall, on scales
dominated by finite infinity regions. The numerical value of $\bH$
is smaller than the global average, $\Hh$, which includes both voids and
bubble walls. Eq.\ (\ref{42}) thus also quantifies the
apparent variance in the Hubble flow below the scale of homogeneity.
Local measurements across single voids of the dominant size, diameter
$30h^{-1}\w{Mpc}$ \citep{HV2},
should give a Hubble ``constant'' which exceeds the global average $\Hm$ by an
amount commensurate to $\Hm-\Hb$. As voids are dominant by volume, an isotropic
average will produce a Hubble ``constant'' {\em greater than} $\Hm$. This
average will steadily decrease from its maximum at $\goesas30h^{-1}$ Mpc
until the scale of homogeneity ($\goesas100h^{-1}\w{Mpc}$) is reached: a
``Hubble bubble'' feature \citep{Tom1,bub1}.

We report the results of three independent cosmological tests. We use
the exact solution \citep{sol} to the Buchert equations
with boundary conditions at the surface of last scattering, $z_i\simeq1100$,
consistent with observations of the CMB. The luminosity distance,
$\dL={\gc}^{-1}\ab\Z0(1+z)\,\rw$, and angular diameter distance,
$\dA=\dL/(1+z)^2$, are referred to the effective dressed geometry
(\ref{wgeom}). We take an initial relative velocity
dispersion, $\hri=0.99999$, between walls and voids, and initial
void volume fraction, $10^{-5}<\fvi<10^{-2}$, at the time of last scattering.
The the results are insensitive to variations of $\hri$ and $\fvi$ for
physically reasonable priors on account of the existence a tracker
solution \citep{sol} to which all solutions tend, to within 1\% by redshifts
of $z\simeq37$. The solutions are then effectively specified
by two independent parameters, which may be taken to be the global
average Hubble constant, $\Hm$, and the present void volume fraction, $\fvn$.

We have tested the luminosity distance of the FB model against the
\citet{Riess06} (Riess07) {\em gold data set} of SneIa and find that for
182 data points and two degrees of freedom the best--fit
$\chi^2=162.7$, i.e., a $\chi^2$ of approximately $0.9$ per degree
of freedom, which is a good fit.
We have performed a Bayesian model comparison of the FB model against a flat
\LCDM\ model with priors $55\le\Hm\le75\kmsMpc$, $0.01\le\OmMn\le0.5$.
This gives a Bayes factor of $\ln B=0.27$ in favour of the FB model, a margin
which is ``not worth more than a bare mention'' \citep{KR} or ``inconclusive''
\citep{Trotta}. Thus the fit of the two models to the Riess07 gold data set
is statistically indistinguishable.

The Riess07 gold data set omits data in
the ``Hubble bubble'' below redshifts of $z\le0.023$. In the \LCDM\ model,
there is no clear theoretical rationale for this; it is merely observed
empirically that a significant reduction in the inferred Hubble constant
occurs at the Hubble bubble scale \citep{bub1}. In the FB model the Hubble
bubble is expected as a feature.

In \figdmu\ we display the residual difference $\DE\mu=
\mu\Ns{FB}-\mu\ns{empty}$, in the standard distance modulus, $\mu=
5\log_{10}(\dL) + 25$, of the best--fit FB model
from that of a coasting Milne universe of the same Hubble constant,
$\Hm=61.7\kmsMpc$, and compare the theoretical curve with binned
data from the Riess07 gold data set. Apparent acceleration occurs
for positive residuals in the range, $z\lsim0.9$. It should be
noted that the exact range of redshifts corresponding to apparent
acceleration also depends on the value of the Hubble constant of
the Milne universe distance modulus used to compute the
residual.
In the FB model the magnitude of the gradient of
the theoretical residual of \figdmu\ by redshift is less than that for
comparable \LCDM\ models. This reflects the fact that the distance modulus
approaches that of a Milne universe at late times.

Statistical confidence limits for the SneIa data are displayed as the
oval contours in the centre of \figcont, in the $(\Hm,\OmMn)$
parameter space. The dressed density parameter is used here, since it
is the one whose numerical value is likely to be closest to that of
a FLRW model, and is thus most familiar. Note that
$\OmMn\simeq\frn12(1-\fvn)
(2+\fvn) 
=\frn18(2+\fvn)^3\OMMn$ \citep{sol}.

In \figcont\ we also overplot parameter ranges for which two
independent cosmological tests have been applied. The first test
is the effective angular diameter of the sound horizon, which very
closely correlates with the angular scale of the first Doppler
peak in the CMB anisotropy spectrum. It is often stated that the angular
position of the first peak is a measure of the spatial curvature of the
universe. However, this deduction relies on the assumption that the
spatial curvature is the same everywhere, appropriate for the FLRW models.
In the present model there are spatial curvature gradients, and we must
revisit the calculation from first principles. Volume--average negative
spatial curvature, which accords with tests of ellipticity
in the CMB anisotropies \citep{ellip1,ellip2}, can nonetheless
be consistent with our local observation of the angular scale of the first
peak \citep{opus}.
\begin{figure}[htb]
\vbox{
\centerline{\scalebox{0.47}
{\includegraphics{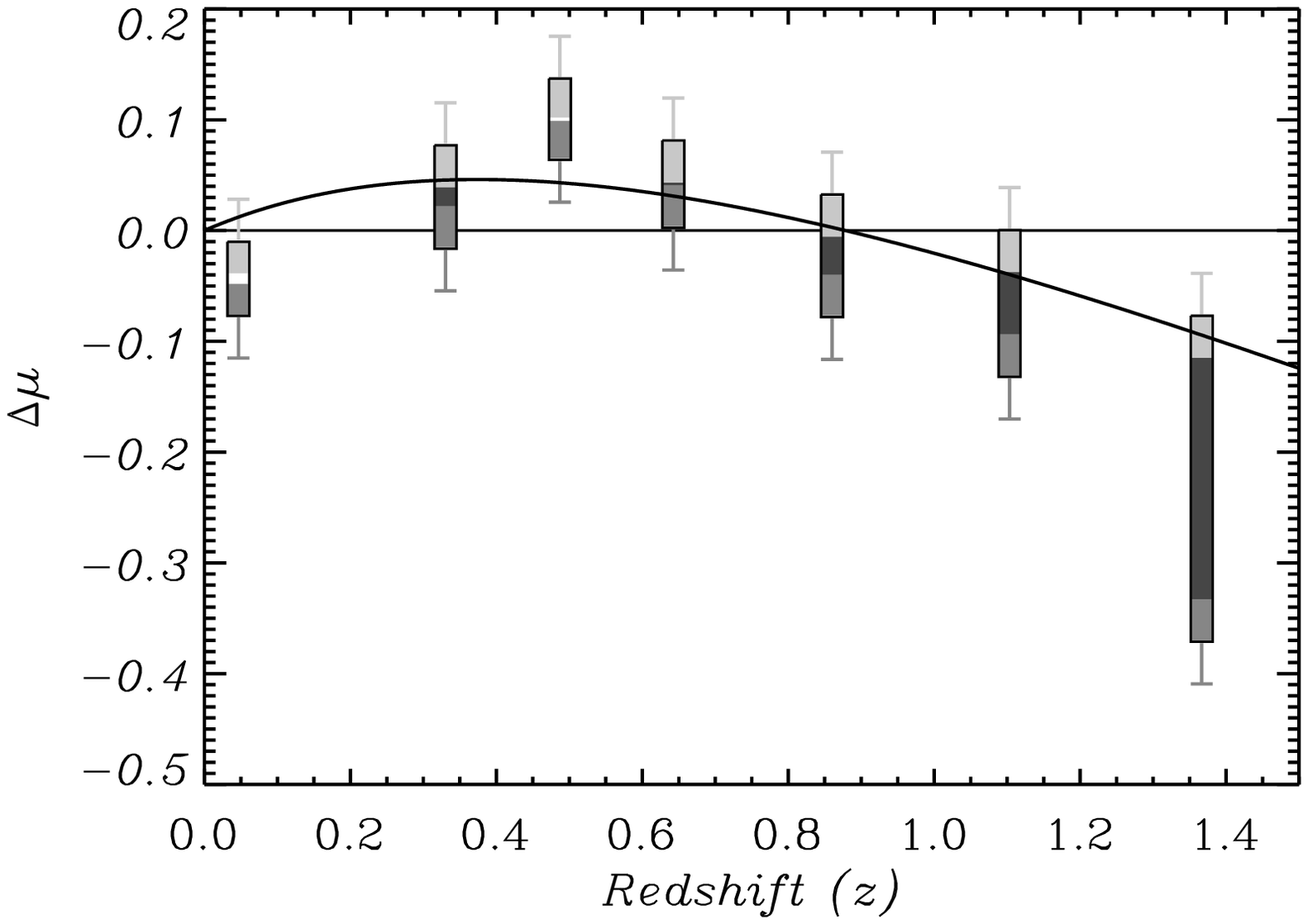}}}
\caption{%
The difference in the distance modulus, $\mu = 5 \log_{10} (\dL)
+ 25$, with $\dL$ in units Mpc, of the FB model with $\Hm=61.7\kmsMpc$,
$\OmMn=0.326$ from that of an empty coasting Milne universe, with the
same value of $\Hm$. The \citet{Riess06} gold data
set of 182 SneIa is binned using the criterion $n_i\DE z_i=5.8$,
where $n$ is the number of data points, and $\DE z_i$ the width
of the $i$th bin. The first bin boundary is set at $z=0.023$ as
``Hubble bubble'' points with $z\le0.023$ are excluded.
Our bins are differ very slightly from those used in Fig.\ 6 of
\citet{Riess06}: 
the single outlier point
at $z=1.755$, falls in its own bin. This point which falls below the
theoretical curve is not shown here, but is included in the $\chi^2$
analysis. We use the original distance
moduli reported at
http://braeburn.pha.jhu.edu/$\goesas$ariess/R06/sn\_sample,
without the suggested systematic subtraction 
of 0.32 mag, as
we follow the Cepheid calibration of \citet{Sandage}. The boxes
show the standard statistical errors for the binned data using the reported
uncertainties,
which already account for luminosity corrections in the MLCS2k2 reduction
\citep{bub1}.
The whiskers indicate how
the residuals move relative to the horizontal axis for the
$2\sigma$ limits on $\Hm$ with $\OmMn=0.326$ fixed: light grey 
corresponds to the 2$\si$ upper bound, and dark grey 
to the 2$\si$ lower bound. The overlap in these two regions has been coloured
black. 
}}
\label{dmu}
\end{figure}
\begin{figure}[htb]
\vbox{
\centerline{\scalebox{0.47}
{\includegraphics{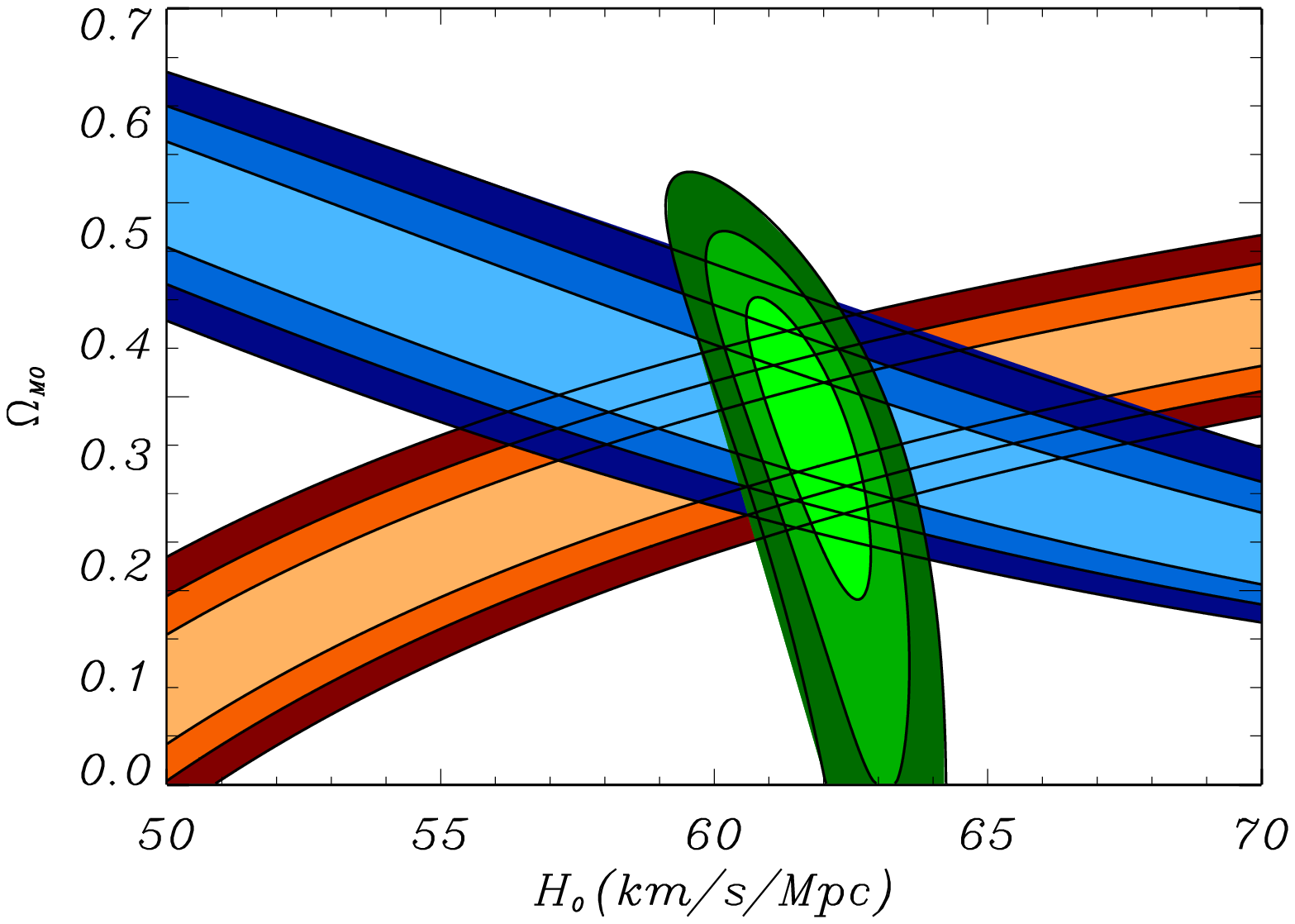}}}
\caption{%
1$\si$, 2$\si$ and 3$\si$ confidence limits (oval contours)
for fits of luminosity distances of type Ia supernovae (SneIa) in the Riess07
gold dataset \citep{Riess06} are compared to parameters within the
($\OmMn$,$\Hm$) plane which fit the angular scale of the sound horizon
$\de=0.01$ rad deduced for WMAP \citep{wmap1,wmap3}, to within 2\%, 4\% and
6\% (contours running top--left to bottom--right); and to parameters which
fit the effective comoving baryon acoustic oscillation (BAO) scale of
$104h^{-1}$Mpc observed in galaxy clustering statistics
\citep{Cole,Eisenstein}, to within 2\%, 4\% and 6\%
(contours running bottom--left to middle--right).}}
\label{contours}
\end{figure}

Ideally we should recompute the spectrum of Doppler peaks for the FB
model. However, this requires considerable effort, as the standard
numerical codes have been written solely for FLRW models, and every step
has to be carefully reconsidered. This task is left for future work. The
test that we apply here is to ask whether parameters exist for
which the effective angular diameter scale of the sound horizon
matches the angular scale of the sound horizon, $\de=0.01$ rad, of the
\LCDM\ model, as determined by WMAP \citep{wmap1}. Since there is no
change to the physics of recombination, but just an overall change
to the {\em calibration} of cosmological parameters, this is
entirely reasonable.

In \figcont\ we plot parameter ranges which match the $\de=0.01$ rad
sound horizon scale to within 2\%, 4\% and 6\%, using the calculation
of the sound horizon given by \citet[\S7.2]{opus}. The 2\% contour
would roughly correspond to the 2$\si$ limit if the WMAP uncertainties for
the \LCDM\ model are maintained. As this can only be confirmed by detailed
computation of the Doppler peaks, the additional levels have been chosen
cautiously. The limits shown have
been arrived at assuming a {\em volume--average} baryon--to--photon ratio
in the range $\etBg=4.6$--$5.6\times10^{-10}$ adopted by
\citet{bbn2} prior to the release of WMAP1. With this range it
is possible to achieve concordance with lithium abundances, while
also better fitting helium abundances. This potentially
resolves an anomaly. With the 2003 WMAP1 release \citep{wmap1}, the
baryon--to--photon ratio was increased to the very upper range of values
that had previously been considered, largely due to the
consequence for the ratio of the heights of the first two Doppler
peaks. This ratio of peak heights is sensitive to the mass
ratio of baryons to non--baryonic dark matter -- rather
than directly to the baryon--to--photon ratio -- as it depends physically on
baryon drag in the primordial plasma. The fit to the Doppler peaks
required more baryons than the range of \citet{bbn2} admitted, when
calibrated with the FLRW model. In the FB calibration, on account of the
difference between the bare and dressed density parameters, a bare value
of $\OMBn\simeq0.03$ nonetheless corresponds to a conventional dressed
value $\Omega\Z{B0}\simeq0.08$, and an overall mass ratio of
baryonic matter to non--baryonic dark matter of
about 1:3, which is larger than for \LCDM.
This would certainly indicate sufficient baryon drag to
accommodate the ratio of the first two peak heights.

The final set of contours plotted in \figcont\ relate to
the independent test of the effective comoving scale of the baryon
acoustic oscillation (BAO), as detected in galaxy clustering
statistics \citep{Cole,Eisenstein}. Similarly to the case of the angular
scale of the sound horizon, given that we do not have the
resources to analyse the galaxy clustering data directly, we begin here with
a simple but effective check. In particular, since the dressed geometry
(\ref{wgeom}) does provide an effective almost--FLRW metric adapted to our
clocks and rods in spatially flat regions,
the effective comoving scale in this dressed geometry should match the
corresponding observed BAO scale of $104h^{-1}$Mpc.
We therefore plot parameter values which
match this scale to within 2\%, 4\% or 6\%.

The best--fit cosmological parameters, using SneIa only, are
$\Hm=61.7^{+1.2}_{-1.1}\kmsMpc$ and $\fvn=0.76^{+0.12}_{-0.09}$, with 1$\si$
uncertainties. The values of the mean lapse function, bare density parameter,
conventional dressed density parameter, mass ratio of non--baryonic dark matter
to baryonic matter, bare Hubble parameter, effective dressed deceleration
parameter and age of the universe measured in a galaxy are respectively:
$\gc=1.381^{+0.061}_{-0.046}$; $\OMMn=0.125^{+0.060}_{-0.069}$;
$\OmMn=0.33^{^+0.11}_{-0.16}$; $(\OMMn-\OMBn)/\OMBn=3.1^{+2.5}_{-2.4}$;
$\Hb=48.2^{+2.0}_{-2.4}\kmsMpc$;
$q=-0.0428^{+0.0120}_{-0.0002}$; $\twn=14.7^{+0.7}_{-0.5}$ Gyr.
Statistical uncertainties from the sound horizon and BAO tests cannot
yet be given, but should significantly reduce the bounds on $\fvn$, $\OmMn$
etc.

One striking feature of \figcont\ is that even if SneIa
are disregarded, the parameters which fit the two independent
tests relating to the sound horizon and the BAO scale agree with each
other, to the accuracy shown, for values of the Hubble constant which include
the value of \citet{Sandage}. However, they do {\em not} agree for the values
of $\Hm$ greater than $70\kmsMpc$ which best--fit the WMAP
data \citep{wmap1,wmap3} with the FLRW model.

The value of the Hubble constant quoted by \citet{Sandage}
has been controversial, given the
14\% difference from values which best--fit the WMAP data with the
\LCDM\ model \citep{wmap1,wmap3}. However, the WMAP analysis only constitutes
a direct measurement of CMB temperature anisotropies; the
determination of cosmological parameters involves model assumptions. We
have removed the assumptions of the FLRW model, in an attempt to model
the universe in terms of the distribution of galaxies that we actually
observe, with an alternative proposal to averaging consistent with
general relativity. Applied to the angular diameter of the sound horizon and
the BAO scale this leads to different cosmological parameters: ones that
agree with the measurement of \citet{Sandage}.

The combination of best--fit cosmological parameters that arises is
particularly interesting. The numerical value of present void volume fraction,
$\fvn$, is identical to that of the dark--energy density
fraction, $\Omega\Z{\Lambda0}$, in the \LCDM\ model with
WMAP \citep{wmap3}. If the FB model is closer to the
correct description of the actual universe, then in trying to fit a FLRW
model, we appear to be led to parameters in which the cosmological constant
is mimicking the effect of voids as far as the WMAP normalization to FLRW
models is concerned. This it does imperfectly, since for a flat \LCDM\ model
$\OmMn=1-\Omega\Z{\Lambda0}$, with the result that the best--fit value
of $\OmMn$ normalized to the CMB does not match the best--fit value of $\OmMn$
for SneIa with the FLRW model, nor for other tests which directly probe
$\OmMn$. For example, it has been recently noted that the values of the
normalization of the primordial spectrum $\si\Z8\goesas0.76$ and matter
content $\OmMn\goesas0.24$ implied by WMAP3 are barely compatible with the
abundances of massive clusters determined from X--ray measurements
\citep{omegam}. For the FB model, by contrast, the dressed density
parameter, $\OmMn$, includes the range preferred in
direct estimations of the conventional matter density parameter.

The integrated Sachs--Wolfe effect provides a further interesting test
to be determined.
Since the observed signal is based on a correlation to clumped structure
\citep{isw}, for large scale averages any difference from the \LCDM\
expectation would largely depend on the difference in expansion history of the
two models. However, we might expect foreground voids to give anisotropies
below the scale of homogeneity, for which evidence is seen 
\citep{bigvoid}.

In this {\em Letter} we have offered preliminary quantitative evidence, via
agreement of independent cosmological tests, that the problem of ``dark
energy'' might be resolved within general relativity. The differences in
cosmological parameters inferred in the \LCDM\ and FB models --
including the average Hubble parameter and its variance, the expansion age,
dressed matter density, baryon--to--photon ratio, baryon--to--dark matter
ratio, CMB ellipticity -- are such that the question as to which provides
the better concordance model can be answered by future observations and new
cosmological tests.\smallskip

\noindent {\em Acknowledgement} This work was supported by the Marsden
Fund of the Royal Society of New Zealand.

\end{document}